\documentclass{moriond}

\def\be{\begin{equation}}
\def\ee{\end{equation}}
\def\bea{\begin{eqnarray}}
\def\eea{\end{eqnarray}}

\newcommand{\Photo}{}

\usepackage{amsmath}
\usepackage{amssymb}
\usepackage{subfigure}

\begin{document}
\vspace*{4cm}
\title{Charm CP violation and searches}

\author{Stefan Schacht}

\address{Department of Physics and Astronomy, University of Manchester,\\ Manchester M13 9PL, United Kingdom}

\maketitle\abstracts{
Charm CP violation is a unique gate to the physics of up-type quarks and allows searches for physics beyond the Standard Model in new and exciting ways, complementary to kaon and $b$~decays. We review recent advances, focusing on symmetry-based methods and providing an outlook on the next challenges at the intensity frontier of charm physics. 
}

\section{Introduction \label{sec:introduction}}

Charm CP violation is a unique gate to explore the flavor structure of up-type quarks and probe for physics 
beyond the Standard Model (BSM) in this sector. In 2019 LHCb discovered charm CP violation by determining the difference of 
CP asymmetries~\cite{LHCb:2019hro,HFLAV:2022esi}
\begin{align}
a_{CP}^{\mathrm{dir}}(D^0\rightarrow K^+K^-) - a_{CP}^{\mathrm{dir}}(D^0\rightarrow \pi^+\pi^-) &= (-0.161\pm 0.028)\%\,, \label{eq:DeltaACP}
\end{align} 
where
\begin{align}
a_{CP}^{\mathrm{dir}}(D^0\rightarrow f) &\equiv \frac{
		\vert A(D^0\rightarrow f)\vert^2 - \vert A(\bar{D}^0\rightarrow f)\vert^2
		}{
		\vert A(D^0\rightarrow f)\vert^2 + \vert A(\bar{D}^0\rightarrow f)\vert^2
		}\,.
\end{align}
The interpretation of Eq.~(\ref{eq:DeltaACP}) and the question if it implies BSM physics or not is an on-going challenge 
for the theory community. 
Direct CP violation is an interference effect of two amplitudes that have a relative weak and strong phase.
In terms of the underlying theory parameters of the two interfering amplitudes, for charm decays we have~\cite{Grossman:2006jg}
\begin{align}
a_{CP}^{\mathrm{dir}} \approx 2 r_{\mathrm{CKM}} r_{\mathrm{QCD}} \sin\varphi_{\mathrm{CKM}} \sin\delta_{\mathrm{QCD}}\,. \label{eq:aCP-theory}
\end{align}
Here, $r_{\mathrm{CKM}}$ is the magnitude of the ratio of the involved Cabibbo-Kobayashi-Maskawa (CKM) matrix elements of the two amplitudes, $r_{\mathrm{QCD}}$ is the 
magnitude of the ratio of the relevant hadronic matrix elements and $\varphi_{\mathrm{CKM}}$ and $\delta_{\mathrm{QCD}}$ are the corresponding relative 
weak and strong phases, respectively. 

In the Standard Model~(SM), the interference effect resulting in $a_{CP}^{\mathrm{dir}}\neq 0$ stems from the different ways one can 
reach the final state of for example two pions from the initial state of a $D^0$. This can proceed either directly via the CKM matrix elements $V_{cd}^* V_{ud}$, or by 
first decaying to an intermediate state that contains strangeness, for example $K^+K^-$, through the coupling $V_{cs}^* V_{us}$, 
which then annihilates and forms a $\pi^+\pi^-$ state:
\begin{align}
D^0 &\overset{V_{cd}^*V_{ud}}{\longrightarrow} \pi^+\pi^-\,,\\
D^0 &\overset{V_{cs}^*V_{us}}{\longrightarrow} K^+ K^-, \dots \overset{\mathrm{QCD}}{\longrightarrow} \pi^+\pi^-\,.
\end{align}
The same happens completely analogously also for $K^+K^-$ final states: 
\begin{align}
D^0 &\overset{V_{cd}^*V_{ud}}{\longrightarrow} \pi^+\pi^-,\dots \overset{\mathrm{QCD} }{\longrightarrow} K^+K^-\,, \\
D^0 &\overset{V_{cs}^*V_{us}}{\longrightarrow} K^+ K^- \,.
\end{align}
The interfering amplitudes have therefore a relative weak phase between the CKM matrix element combinations $V_{cd}^* V_{ud}$ and  $V_{cs}^* V_{us}$ as well as a relative strong phase from the involved non-perturbative QCD effects.

The fact that we know the involved CKM matrix elements from the global CKM fit,~\emph{i.e.},~from kaon and $b$ decays, leads to the SM estimate
\begin{align}
\Delta a_{CP}^{\mathrm{dir}} \sim 10^{-3} \times r_{\mathrm{QCD}} \sin\delta_{\mathrm{QCD}}\,.
\end{align}
However, we are as of now not able to reliably calculate the underlying hadronic physics from first principles, so we do not know \emph{a priori} the 
size of $r_{\mathrm{QCD}}$ and $\delta_{\mathrm{QCD}}$. The ratio $r_{\mathrm{QCD}}$ is also known as the \lq\lq{}penguin over tree\rq\rq{} ratio.
Note however, that this commonly used term denotes more precisely just the ratio of CKM-subleading over CKM-leading amplitudes. The penguin diagrams are actually penguin contractions of tree operators~\cite{Brod:2011re,Brod:2012ud}, and the CKM-leading amplitudes include also other topologies 
than tree diagrams, see,~\emph{e.g.},~Ref.~\cite{Muller:2015lua}.

In terms of $U$-spin matrix elements, $r_{\mathrm{QCD}}$ is given as the ratio of $\Delta U=0$ over $\Delta U=1$ hadronic matrix elements.
$U$-spin is a SU(2) subgroup of the approximate SU(3)-flavor symmetry of the QCD Lagrangian under 
unitary rotations of the light quarks $u$, $d$ and $s$. $U$-spin is the symmetry group which connects $d$ and $s$~quarks.
Another such subgroup is isospin, which connects $u$ and $d$~quarks.

Assuming the SM and $\delta_{\mathrm{QCD}} = \mathcal{O}(1)$ from large rescattering effects, Eq.~(\ref{eq:DeltaACP}) implies~\cite{Grossman:2019xcj}
\begin{align}
r_{\mathrm{QCD}}^{\mathrm{Exp}} &= \mathcal{O}(1)\,. \label{eq:rQCDExp}
\end{align} 
The key question in order to answer the question whether Eq.~(\ref{eq:DeltaACP}) implies BSM physics or not is therefore equivalent to the 
question for the size of $r_{\mathrm{QCD}}$ in the SM.
However, in hadronic charm decays, it is not clear if methods known from kaon and $b$ physics work.
The key issue is if we can overcome soft QCD in charm decays.
At the relevant scale, the strong coupling $\alpha_s$ is too large in order to make a meaningful expansion.
Also, in general an expansion in $\Lambda_{\mathrm{QCD}}/m_c$ does not work. Note that in some cases this might be different, like for the charm hadron lifetimes, where heavy quark methods show promising results~\cite{King:2021xqp,Gratrex:2022xpm}. For hadronic decays we are however in a different position. For these, we need to work with a smaller toolbox and find new strategies in order to interpret the data. 

Eq.~(\ref{eq:rQCDExp}) is consistent with large $\mathcal{O}(1)$ non-perturbative effects from low energy QCD, which result in~\cite{Brod:2011re,Grossman:2019xcj,Schacht:2021jaz}
\begin{align}
r_{\mathrm{QCD}}^{\mathrm{SM}} = \mathcal{O}(1)\,,
\end{align}
in agreement with expectations based on enhancements that we also see in the ratio of $\Delta I=1/2$ over $\Delta I=3/2$ $D\rightarrow \pi\pi$ isospin matrix elements~\cite{Franco:2012ck}, see Ref.~\cite{Gavrilova:2023fzy} for a recent update. 
On the other hand, works based on light cone sum rules (LCSR) predict~\cite{Khodjamirian:2017zdu,Chala:2019fdb,Lenz:2023rlq}
\begin{align}
r_{\mathrm{QCD}}^{\mathrm{SM}} = \mathcal{O}(0.1)\,.
\end{align}
Strikingly, this prediction is one order of magnitude below the experimental determination in Eq.~(\ref{eq:rQCDExp}).
Another promising ansatz to treat hadronic charm decays makes use of the available $\pi\pi$/$KK$ rescattering data, see Refs.~\cite{Franco:2012ck,Bediaga:2022sxw,Pich:2023kim} for detailed discussions. 
First ideas on the lattice can be found in Ref.~\cite{Hansen:2012tf}, however the amount of intermediate states proves challenging.

In the following, we review some of the recent developments. In Sec.~\ref{sec:isospin}, we explain the methodology for the extraction of the 
penguin over tree ratio from data with theory uncertainties of $\sim1\%$.
In Sec.~\ref{sec:uspin} we discuss how recent measurements of CP asymmetries of singly-Cabibbo suppressed decays allow to probe 
the breaking of the $U$-spin symmetry in CKM-subleading amplitudes.
In Sec.~\ref{sec:higher-order} we summarize recent insights on sum rules at higher order in the $U$-spin expansion.
We conclude in Sec.~\ref{sec:conclusions}.

\section{Penguin over tree extraction with $\sim1\%$ theory uncertainty \label{sec:isospin}}

As mentioned above, the extraction of the penguin over tree ratio Eq.~(\ref{eq:rQCDExp}) from $\Delta a_{CP}^{\mathrm{dir}}$ depends on the  
assumption that the relevant strong phase is $\mathcal{O}(1)$, see,~\emph{e.g.},~Ref.~\cite{Grossman:2019xcj}.
For a precision extraction of $r_{\mathrm{QCD}}$ it would of course be advantageous not to rely on such an assumption. 
It is known that the strong phase between the interfering amplitudes can be obtained in principle from measurements of 
time-dependent CP violation or quantum-correlated decays~\cite{Grossman:2019xcj,Xing:1996pn,Gronau:2001nr,Bevan:2011up,Bevan:2013xla,Grossman:2012eb,Xing:2019uzz,Schacht:2022kuj}.
A difficulty in charm decays is the very slow oscillation between the mesons.
Unlike the time-dependence of $B$ decays, in charm decays we aim just for the linear term of the time dependence, $\Delta Y^f$, in
\begin{align} 
A_{CP}(f,t) \approx a_{CP}^f + \Delta Y^f \frac{t}{\tau_{D^0}}\,.
\end{align}
Current determinations of $\Delta Y^f$ are compatible with zero~\cite{LHCb:2022lry,LHCb:2024jpt}.
The slopes $\Delta Y^f$ of different decay channels with final states $f$ have a universal and a non-universal contribution
and the splitting of $\Delta Y^f$ between $D^0\rightarrow K^+K^-$ and $D^0\rightarrow \pi^+\pi^-$ is
formally suppressed at the second power of $U$-spin breaking effects~\cite{Kagan:2020vri}.

The case of $D^0\rightarrow \pi^+\pi^-$ is especially interesting as here we currently see hints for very large CP violation~\cite{LHCb:2022lry}.
In the $D\rightarrow \pi\pi$ system we have the additional advantage of isospin symmetry relating the observables of $D^0\rightarrow \pi^+\pi^-$, 
$D^0\rightarrow \pi^0\pi^0$ and $D^0\rightarrow \pi^+\pi^0$. Isospin breaking effects are expected only at the $1\%$ level. 
Assuming the SM, isospin allows the determination of the relative strong phase between penguin and tree amplitudes from direct CP asymmetries and branching ratios only. Corresponding closed-form expressions have been derived recently in Ref.~\cite{Gavrilova:2023fzy}.
$D\rightarrow \pi\pi$ has the same underlying group-theory structure as $B\rightarrow \pi\pi$~\cite{Gronau:1990ka}, however,
the different approximations that apply to charm and $B$ decays result in different implications for how 
the underlying theory parameters relate to direct CP asymmetries and branching ratios.
For example, an important feature of charm decays is that the physics of branching ratios is to a very good approximation only governed 
by the CKM-leading amplitude because the CKM-subleading amplitude is strongly suppressed. 
This leads to a decoupling of the physics of branching ratios and CP asymmetries.

An interesting feature of $D\rightarrow \pi\pi$ is that due to the underlying isospin symmetry, knowledge about the penguin-over-tree ratio in one decay channel translates into a prediction about this ratio in the other one, namely~\cite{Gavrilova:2023fzy}
\begin{align}
\frac{
r^{00}
}{
r^{+-}
} &= \sqrt{
\frac{1}{2} \frac{\mathcal{B}(D^0\rightarrow \pi^+\pi^-)}{\mathcal{B}(D^0\rightarrow \pi^0\pi^0)}
 \frac{\mathcal{P}^{00}}{\mathcal{P}^{+-} }
}\,,
\end{align}
where $\mathcal{P}^f$ are the corresponding phase space factors and $r^{00}$ and $r^{+-}$ are the penguin over tree ratios of the isospin decomposition of  $D^0\rightarrow \pi^0\pi^0$ and $D^0\rightarrow \pi^+\pi^-$ decays, respectively.
With current data it is found~\cite{Gavrilova:2023fzy}
\begin{align}
r^{+-} &=  5.5^{+14.2}_{-2.7} \,,\\
r^{00} &=  5.2^{+13.3}_{-2.4} \,,
\end{align}
\emph{i.e.},~the central values are unexpectedly large. With future data from LHCb Upgrade~II and Belle~II it will be possible to reduce the uncertainties considerably~\cite{Gavrilova:2023fzy}.
Once also time-dependent CP violation measurements are available, the above methodology can be used in order to not only determine, but overconstrain the 
$D\rightarrow \pi\pi$ system and thereby probe for BSM physics.

\section{Testing the $U$-spin symmetry in CKM-subleading amplitudes \label{sec:uspin}}

Recent new measurements by LHCb went beyond just the measurement of the combination of CP asymmetries in Eq.~(\ref{eq:DeltaACP}) and provide
separate determinations of the involved CP asymmetries~\cite{LHCb:2022lry}
\begin{align}
a_{CP}^{\mathrm{dir}}(D^0\rightarrow K^+K^-)      &= (7.7\pm 5.7)\cdot 10^{-4}\,,  \label{eq:aCPKK}\\
a_{CP}^{\mathrm{dir}}(D^0\rightarrow \pi^+\pi^- ) &= ( 23.2\pm 6.1) \cdot 10^{-4}\,, \label{eq:aCPpipi}
\end{align}
implying the first evidence for CP violation in a single decay channel.
The central values of Eqs.~(\ref{eq:aCPKK}) and (\ref{eq:aCPpipi}) violate the $U$-spin limit sum rule~\cite{Grossman:2012ry}
\begin{align}
a_{CP}^{\mathrm{dir}}(D^0\rightarrow K^+K^-) +
a_{CP}^{\mathrm{dir}}(D^0\rightarrow \pi^+\pi^- ) &= 0\,. \label{eq:U-spin}
\end{align}
While a violation of Eq.~(\ref{eq:U-spin}) is expected, the amount goes beyond the generic expectation of $\sim 30\%$ SU(3)$_F$ breaking, however 
yet with large errors. 
With future data, additional sum rules such as~\cite{Grossman:2012ry} 
\begin{align}
a_{CP}^{\mathrm{dir}}(D_s^+\rightarrow K^0\pi^+) + 
a_{CP}^{\mathrm{dir}}(D^+\rightarrow \overline{K}^0 K^+) = 0 \label{eq:U-spin-sum-rule-2} 
\end{align}
can be tested in order to study if a consistent picture emerges~\cite{Schacht:2022kuj}.
Improved versions of the sum rules Eqs.~(\ref{eq:U-spin}) and (\ref{eq:U-spin-sum-rule-2}) connect the decays
\begin{align}
& a_{CP}^{\mathrm{dir}}(D^0\rightarrow K^+K^-)\,, \quad
a_{CP}^{\mathrm{dir}}(D^0\rightarrow \pi^+\pi^-)\,,\quad
a_{CP}^{\mathrm{dir}}(D^0\rightarrow \pi^0\pi^0)\,, \quad \text{and}\nonumber\\ 
& a_{CP}^{\mathrm{dir}}(D^+\rightarrow K_SK^+)\,,\quad
a_{CP}^{\mathrm{dir}}(D_s^+\rightarrow K_S\pi^+)\,,\quad
a_{CP}^{\mathrm{dir}}(D_s^+\rightarrow K^+\pi^0)\,,\nonumber
\end{align}
respectively, and can be found in Ref.~\cite{Muller:2015rna}. The latter highlight the synergy and complementarity of LHCb~\cite{LHCb:2019hro,LHCb:2021rou,LHCb:2022lry} and Belle/Belle~II~\cite{Belle:2014evd,Belle:2021ygw}. 
Explanations of large $U$-spin breaking in the CKM-subleading amplitudes also include $Z'$ models~\cite{Bause:2020obd,Bause:2022jes},
as these generate operators with the flavor content $\bar{s}c\bar{u}s$ and $\bar{d}c\bar{u}d$ with non-universal coefficients.
Ref.~\cite{Bause:2022jes} presents viable models with a leptophobic $Z'$ below $O(20\,\text{GeV})$, including a pattern of CP violation
in $D\rightarrow \pi\pi$ that includes a violation of the isospin sum rule~\cite{Grossman:2012eb,Buccella:1992sg}
\begin{align}
a_{CP}^{\mathrm{dir}}(D^+\rightarrow\pi^+\pi^0) &= 0\,.
\end{align}
More study is needed of the puzzling violation of Eq.~(\ref{eq:U-spin}) once more data becomes available.

\section{Solving the Problem of Higher Order $U$-Spin \label{sec:higher-order}} 

The discussion in Sec.~\ref{sec:uspin} motivates to go to higher order in the $U$-spin expansion in order to derive sum rules 
that hold to higher precision than Eq.~(\ref{eq:U-spin}). 
Going to higher order in SU(3)$_F$ breaking is complicated by the proliferation of parameters, visibly already at
next to leading order~\cite{Hiller:2012xm}.
In Ref.~\cite{Gavrilova:2022hbx} we derived new theorems which enable calculations to arbitrary order in the $U$-spin expansion on the amplitude level.
This allows especially also to determine \emph{a priori} up to which order sum rules actually exist. The new method uses symmetry properties of Clebsch-Gordan coefficients and the fact that any given system can be reduced to the discussion of a doublet-only system out of which the given system is constructed.

Many more opportunities lie ahead in this direction. We expect that this methodology will be especially useful for multi-body decays.
Precision sum rules at high order are likely for these to exist due to the higher combinatorial possibilities in comparison to two-body decays. 

The first step regarding three-body decays however will be to probe the ratio of $\Delta U=0$ over $\Delta U=1$ matrix elements of underlying pseudo two-body $D\rightarrow VP$ decays in order to see if their order of magnitude agrees with the analogous ratio for $D\rightarrow PP$ decays in Eq.~(\ref{eq:rQCDExp})~\cite{Dery:2021mll}. This enables an important consistency check of the hierarchies of $U$-spin matrix elements.
Such endeavors are complementary to model-independent searches for CP violation in multi-body decays for example with the energy test~\cite{Parkes:2016yie,LHCb:2023mwc,LHCb:2023rae}. Another frontier are charmed baryon decays, see recently Refs.~\cite{LHCb:2017hwf,Belle:2022uod,Grossman:2018ptn,Jia:2019zxi}, to which the methodology of Ref.~\cite{Gavrilova:2022hbx} also applies.

\section{Conclusions \label{sec:conclusions}}

This is just the beginning of the exploration of charm CP violation.
Until recently we had only a measurement of $\Delta a_{CP}^{\mathrm{dir}}$. Now, with the evidence of CP violation
in $D^0\rightarrow \pi^+\pi^-$ we have two data points. In order to benefit from symmetry-based methods for the 
interpretation of the data, we need more measurements of CP asymmetries of singly-Cabibbo suppressed decays.
Only then we will be able to test the corresponding sum rules.
Time-dependent measurements will be very important to use isospin methods to overconstrain the $D\rightarrow \pi\pi$ system.
The development of new higher order sum rules will provide opportunities to benefit from the abundant data on  
multi-body decays. 
The key question if we can tell a loop from a tree remains a challenge for charm theory, but 
no matter what, we will learn something new: about QCD, and maybe also about new physics. 

\section*{Acknowledgments}

S.S. is supported by a Stephen Hawking Fellowship from UKRI under reference EP/T01623X/1 and the STFC research grants ST/T001038/1 and ST/X00077X/1.

\section*{References}

\begin{scriptsize}
\bibliography{schacht}

\begin{thebibliography}{10}

\bibitem{LHCb:2019hro}
Roel Aaij et~al.
\newblock {Observation of CP Violation in Charm Decays}.
\newblock {\em Phys. Rev. Lett.}, 122(21):211803, 2019.

\bibitem{HFLAV:2022esi}
Yasmine~Sara Amhis et~al.
\newblock {Averages of b-hadron, c-hadron, and \ensuremath{\tau}-lepton
  properties as of 2021}.
\newblock {\em Phys. Rev. D}, 107(5):052008, 2023.

\bibitem{Grossman:2006jg}
Yuval Grossman, Alexander~L. Kagan, and Yosef Nir.
\newblock {New physics and CP violation in singly Cabibbo suppressed D decays}.
\newblock {\em Phys. Rev. D}, 75:036008, 2007.

\bibitem{Brod:2011re}
Joachim Brod, Alexander~L. Kagan, and Jure Zupan.
\newblock {Size of direct CP violation in singly Cabibbo-suppressed D decays}.
\newblock {\em Phys. Rev. D}, 86:014023, 2012.

\bibitem{Brod:2012ud}
Joachim Brod, Yuval Grossman, Alexander~L. Kagan, and Jure Zupan.
\newblock {A Consistent Picture for Large Penguins in D -\ensuremath{>} pi+
  pi-, K+ K-}.
\newblock {\em JHEP}, 10:161, 2012.

\bibitem{Muller:2015lua}
Sarah M\"uller, Ulrich Nierste, and Stefan Schacht.
\newblock {Topological amplitudes in $D$ decays to two pseudoscalars: A global
  analysis with linear $SU(3)_F$ breaking}.
\newblock {\em Phys. Rev. D}, 92(1):014004, 2015.

\bibitem{Grossman:2019xcj}
Yuval Grossman and Stefan Schacht.
\newblock {The emergence of the $\Delta U=0$ rule in charm physics}.
\newblock {\em JHEP}, 07:020, 2019.

\bibitem{King:2021xqp}
Daniel King, Alexander Lenz, Maria~Laura Piscopo, Thomas Rauh, Aleksey~V.
  Rusov, and Christos Vlahos.
\newblock {Revisiting inclusive decay widths of charmed mesons}.
\newblock {\em JHEP}, 08:241, 2022.

\bibitem{Gratrex:2022xpm}
James Gratrex, Bla\v{z}enka Meli\'c, and Ivan Ni\v{s}and\v{z}i\'c.
\newblock {Lifetimes of singly charmed hadrons}.
\newblock {\em JHEP}, 07:058, 2022.

\bibitem{Schacht:2021jaz}
Stefan Schacht and Amarjit Soni.
\newblock {Enhancement of charm CP violation due to nearby resonances}.
\newblock {\em Phys. Lett. B}, 825:136855, 2022.

\bibitem{Franco:2012ck}
Enrico Franco, Satoshi Mishima, and Luca Silvestrini.
\newblock {The Standard Model confronts CP violation in $D^0 \to \pi^+\pi^-$
  and $D^0 \to K^+K^-$}.
\newblock {\em JHEP}, 05:140, 2012.

\bibitem{Gavrilova:2023fzy}
Margarita Gavrilova, Yuval Grossman, and Stefan Schacht.
\newblock {Determination of the
  D\textrightarrow{}\ensuremath{\pi}\ensuremath{\pi} ratio of penguin over tree
  diagrams}.
\newblock {\em Phys. Rev. D}, 109(3):033011, 2024.

\bibitem{Khodjamirian:2017zdu}
Alexander Khodjamirian and Alexey~A. Petrov.
\newblock {Direct CP asymmetry in $D\to \pi^-\pi^+$ and $D\to K^-K^+$ in
  QCD-based approach}.
\newblock {\em Phys. Lett. B}, 774:235--242, 2017.

\bibitem{Chala:2019fdb}
Mikael Chala, Alexander Lenz, Aleksey~V. Rusov, and Jakub Scholtz.
\newblock {$\Delta A_{CP}$ within the Standard Model and beyond}.
\newblock {\em JHEP}, 07:161, 2019.

\bibitem{Lenz:2023rlq}
Alexander Lenz, Maria~Laura Piscopo, and Aleksey~V. Rusov.
\newblock {Two body non-leptonic D$^{0}$ decays from LCSR and implications
  for${\Delta a}_{{\text{CP}}}^{{\text{dir}}}$}.
\newblock {\em JHEP}, 03:151, 2024.

\bibitem{Bediaga:2022sxw}
Ignacio Bediaga, Tobias Frederico, and Patricia~C. Magalh\~aes.
\newblock {Enhanced Charm CP Asymmetries from Final State Interactions}.
\newblock {\em Phys. Rev. Lett.}, 131(5):051802, 2023.

\bibitem{Pich:2023kim}
Antonio Pich, Eleftheria Solomonidi, and Luiz Vale~Silva.
\newblock {Final-state interactions in the CP asymmetries of charm-meson
  two-body decays}.
\newblock {\em Phys. Rev. D}, 108(3):036026, 2023.

\bibitem{Hansen:2012tf}
Maxwell~T. Hansen and Stephen~R. Sharpe.
\newblock {Multiple-channel generalization of Lellouch-Luscher formula}.
\newblock {\em Phys. Rev. D}, 86:016007, 2012.

\bibitem{Xing:1996pn}
Zhi-zhong Xing.
\newblock {D0 - anti-D0 mixing and CP violation in neutral D meson decays}.
\newblock {\em Phys. Rev. D}, 55:196--218, 1997.

\bibitem{Gronau:2001nr}
Michael Gronau, Yuval Grossman, and Jonathan~L. Rosner.
\newblock {Measuring D0 - anti-D0 mixing and relative strong phases at a charm
  factory}.
\newblock {\em Phys. Lett. B}, 508:37--43, 2001.

\bibitem{Bevan:2011up}
A.~J. Bevan, G.~Inguglia, and B.~Meadows.
\newblock {Time-dependent $CP$ asymmetries in $D$ and $B$ decays}.
\newblock {\em Phys. Rev. D}, 84(11):114009, 2011.
\newblock [Erratum: Phys.Rev.D 87, 039905 (2013)].

\bibitem{Bevan:2013xla}
A.~J. Bevan and B.~T. Meadows.
\newblock {Bounding hadronic uncertainties in $c\to u$ decays}.
\newblock {\em Phys. Rev. D}, 90(9):094028, 2014.

\bibitem{Grossman:2012eb}
Yuval Grossman, Alexander~L. Kagan, and Jure Zupan.
\newblock {Testing for new physics in singly Cabibbo suppressed D decays}.
\newblock {\em Phys. Rev. D}, 85:114036, 2012.

\bibitem{Xing:2019uzz}
Zhi-zhong Xing.
\newblock {A U-spin prediction for the CP-forbidden transition $e^+ e^- \to
  D^0\bar{D}^0 \to (K^+ K^-)_D (\pi^+ \pi^-)_D$}.
\newblock {\em Mod. Phys. Lett. A}, 34(29):1950238, 2019.

\bibitem{Schacht:2022kuj}
Stefan Schacht.
\newblock {A U-spin anomaly in charm CP violation}.
\newblock {\em JHEP}, 03:205, 2023.

\bibitem{LHCb:2022lry}
R.~Aaij et~al.
\newblock {Measurement of the Time-Integrated CP Asymmetry in
  D0\textrightarrow{}K-K+ Decays}.
\newblock {\em Phys. Rev. Lett.}, 131(9):091802, 2023.

\bibitem{LHCb:2024jpt}
Roel Aaij et~al.
\newblock {Search for time-dependent $CP$ violation in $D^0 \rightarrow \pi^+
  \pi^- \pi^0$ decays}.
\newblock 5 2024.

\bibitem{Kagan:2020vri}
Alexander~L. Kagan and Luca Silvestrini.
\newblock {Dispersive and absorptive $CP$ violation in $D^0- \overline{D^0}$
  mixing}.
\newblock {\em Phys. Rev. D}, 103(5):053008, 2021.

\bibitem{Gronau:1990ka}
Michael Gronau and David London.
\newblock {Isospin analysis of CP asymmetries in B decays}.
\newblock {\em Phys. Rev. Lett.}, 65:3381--3384, 1990.

\bibitem{Grossman:2012ry}
Yuval Grossman and Dean~J. Robinson.
\newblock {SU(3) Sum Rules for Charm Decay}.
\newblock {\em JHEP}, 04:067, 2013.

\bibitem{Muller:2015rna}
Sarah M\"uller, Ulrich Nierste, and Stefan Schacht.
\newblock {Sum Rules of Charm CP Asymmetries beyond the SU(3)$_F$ Limit}.
\newblock {\em Phys. Rev. Lett.}, 115(25):251802, 2015.

\bibitem{LHCb:2021rou}
Roel Aaij et~al.
\newblock {Search for CP violation in $ {D}_{(s)}^{+}\to {h}^{+}{\pi}^0 $ and $
  {D}_{(s)}^{+}\to {h}^{+}\eta $ decays}.
\newblock {\em JHEP}, 06:019, 2021.

\bibitem{Belle:2014evd}
N.~K. Nisar et~al.
\newblock {Search for {CP} violation in $D^0 \to \pi^0 \pi^0$ decays}.
\newblock {\em Phys. Rev. Lett.}, 112:211601, 2014.

\bibitem{Belle:2021ygw}
Y.~Guan et~al.
\newblock {Measurement of branching fractions and $CP$ asymmetries for $D_s^{+}
  \rightarrow K^+ (\eta, \pi^0) $ and $D_s^{+} \rightarrow \pi^+ (\eta, \pi^0)$
  decays at Belle}.
\newblock {\em Phys. Rev. D}, 103:112005, 2021.

\bibitem{Bause:2020obd}
Rigo Bause, Hector Gisbert, Marcel Golz, and Gudrun Hiller.
\newblock {Exploiting $CP$-asymmetries in rare charm decays}.
\newblock {\em Phys. Rev. D}, 101(11):115006, 2020.

\bibitem{Bause:2022jes}
Rigo Bause, Hector Gisbert, Gudrun Hiller, Tim H\"ohne, Daniel~F. Litim, and
  Tom Steudtner.
\newblock {U-spin-CP anomaly in charm}.
\newblock {\em Phys. Rev. D}, 108(3):035005, 2023.

\bibitem{Buccella:1992sg}
F.~Buccella, Maurizio Lusignoli, G.~Mangano, G.~Miele, A.~Pugliese, and Pietro
  Santorelli.
\newblock {CP Violating asymmetries in charged D meson decays}.
\newblock {\em Phys. Lett. B}, 302:319--325, 1993.

\bibitem{Hiller:2012xm}
Gudrun Hiller, Martin Jung, and Stefan Schacht.
\newblock {SU(3)-flavor anatomy of nonleptonic charm decays}.
\newblock {\em Phys. Rev. D}, 87(1):014024, 2013.

\bibitem{Gavrilova:2022hbx}
Margarita Gavrilova, Yuval Grossman, and Stefan Schacht.
\newblock {The mathematical structure of U-spin amplitude sum rules}.
\newblock {\em JHEP}, 08:278, 2022.

\bibitem{Dery:2021mll}
Avital Dery, Yuval Grossman, Stefan Schacht, and Abner Soffer.
\newblock {Probing the $\Delta U=0$ rule in three body charm decays}.
\newblock {\em JHEP}, 05:179, 2021.

\bibitem{Parkes:2016yie}
Chris Parkes, Shanzhen Chen, Jolanta Brodzicka, Marco Gersabeck, Giulio Dujany,
  and William Barter.
\newblock {On model-independent searches for direct CP violation in multi-body
  decays}.
\newblock {\em J. Phys. G}, 44(8):085001, 2017.

\bibitem{LHCb:2023mwc}
R.~Aaij et~al.
\newblock {Search for CP violation in the phase space of
  D$^{0}$\textrightarrow{}
  \ensuremath{\pi}$^{-}$\ensuremath{\pi}$^{+}$\ensuremath{\pi}$^{0}$ decays
  with the energy test}.
\newblock {\em JHEP}, 09:129, 2023.
\newblock [Erratum: JHEP 04, 040 (2024)].

\bibitem{LHCb:2023rae}
Roel Aaij et~al.
\newblock {Search for CP violation in the phase space of $ {D}^0\to
  {K}_S^0{K}^{\pm }{\pi}^{\mp } $ decays with the energy test}.
\newblock {\em JHEP}, 03:107, 2024.

\bibitem{LHCb:2017hwf}
R.~Aaij et~al.
\newblock {A measurement of the $CP$ asymmetry difference in
  $\varLambda_{c}^{+} \to pK^{-}K^{+}$ and $p\pi^{-}\pi^{+}$ decays}.
\newblock {\em JHEP}, 03:182, 2018.

\bibitem{Belle:2022uod}
L.~K. Li et~al.
\newblock {Search for CP violation and measurement of branching fractions and
  decay asymmetry parameters for
  \ensuremath{\Lambda}c+\textrightarrow{}\ensuremath{\Lambda}h+ and
  \ensuremath{\Lambda}c+\textrightarrow{}\ensuremath{\Sigma}0h+
  (h=K,\ensuremath{\pi})}.
\newblock {\em Sci. Bull.}, 68:583--592, 2023.

\bibitem{Grossman:2018ptn}
Yuval Grossman and Stefan Schacht.
\newblock {U-Spin Sum Rules for CP Asymmetries of Three-Body Charmed Baryon
  Decays}.
\newblock {\em Phys. Rev. D}, 99(3):033005, 2019.

\bibitem{Jia:2019zxi}
Cai-Ping Jia, Di~Wang, and Fu-Sheng Yu.
\newblock {Charmed baryon decays in $SU(3)_F$ symmetry}.
\newblock {\em Nucl. Phys. B}, 956:115048, 2020.

\end{thebibliography}
\end{scriptsize}

\end{document}